# THE UNUSUAL VOLATILE COMPOSITION OF THE HALLEY-TYPE COMET 8P/TUTTLE: ADDRESSING THE EXISTENCE OF AN INNER OORT CLOUD[+]

H. Böhnhardt[1], M. J. Mumma[2*], G. L. Villanueva[2], M. A. DiSanti[2], B. P. Bonev[2,3], M. Lippi[1], & H. U. Käufl[4]

[1] Max-Planck Institute for Solar System Research, Max-Planck-Str. 2, D-37191 Katlenburg-Lindau, Germany

[2] Solar System Exploration Division, MS 690.3, NASA Goddard Space Flight Center, Greenbelt, MD 20771, USA

[3] Dept. of Physics, The Catholic University of America, Washington, D. C., USA

[4] European Southern Observatory, Karl-Schwarzschild-Str. 2, D-85748 Garching bei München, Germany

*To whom correspondence should be addressed.  E-mail: michael.j.mumma@nasa.gov

**Abstract:**

We measured organic volatiles ($CH_4$, $CH_3OH$, $C_2H_6$, $H_2CO$), CO, and water in comet 8P/Tuttle, a comet from the Oort cloud reservoir now in a short-period Halley-type orbit.  We compare its composition with two other comets in Halley-type orbits, and with comets of the "organics-normal" and "organics-depleted" classes.  Chemical gradients are expected in the comet-forming region of the proto-planetary disk, and an individual comet should reflect its specific heritage.  If Halley-type comets came from the inner Oort cloud as proposed, we see no common characteristics that could distinguish such comets from those that were stored in the outer Oort cloud.

Subject Headings:  Comets, organics, composition, Halley-type

___________________________________________________________________________

The compositions of cometary nuclei provide key evidence for understanding the formation and evolution of matter in the early solar system, and for assessing delivery of water and organics to early Earth. Chemical gradients are expected in the comet-forming region of the proto-planetary disk, and an individual comet should reflect its specific heritage. Once ejected to distant reservoirs, cosmic rays can process the cometary nucleus to a depth of several meters while material at greater depth remains largely unaltered. The processed layer is lost (while active) on the first few passages through the inner planetary region, and thereafter the comet releases primitive material that preserves a record of its formative conditions (Mumma et al. 1993). A classification based on direct measurement of parent volatiles (i.e., native ices) is required.

Comet 1P/Halley was the first comet classified by both ground-based and *in situ* studies of its parent volatiles, and it revealed high deuterium content and an organic composition similar to that of dense interstellar clouds. This stimulated vigorous development of ground-based search strategies and observational measurements (molecular spectroscopy) of parent volatiles among comets, now revealing (at least) two distinct classes based on their organic composition ("normal" and depleted) (Crovisier 2007, Bockelée-Morvan et al. 2004, Mumma et al. 2003). The most recent development is the deployment of a powerful new instrument (CRIRES at the European Southern Observatory) and its application to comets. Here, we report the composition of comet 8P/Tuttle, measured with CRIRES in January 2008, and we compare it with that of Halley and other comets.

Comets today reside in two distinct reservoirs – the Oort cloud and the Kuiper-belt that in turn is sub-divided into the classical Kuiper-belt, the scattered disk, and the detached (or extended disk) population (Gladman 2005). Comets injected into the inner planetary system are grouped dynamically into several orbital categories: nearly isotropic (subgroups: dynamically-new, long-period - hereafter LPC, Halley-type - hereafter HTC), and ecliptic (subgroups: Centaur-type, Encke-type, and Jupiter-family comets - hereafter JFCs). Ecliptic comets come from the Kuiper belt reservoir while the nearly isotropic comets come from the Oort cloud (Levison 1996).

It once was thought that most JFCs formed in the Kuiper-belt region ($R_h > 30$ AU) while Oort-cloud comets (OCs) formed in the giant-planets' region (5-30 AU) of the protoplanetary disk. Following this paradigm, the OCs formed in a warmer environment than the JFCs, thus, some differences in chemistry or dust properties should be found between JFCs and OCs but each reservoir would contain distinct chemical populations. A new paradigm has been proposed (the "Nice" model; Gomes et al. 2005, Morbidelli et al. 2008) and it predicts that comets formed in the outer proto-planetary disk entered both the Oort cloud and the Kuiper disk, though likely in different proportions. Thus, a comparison of chemical taxonomies in JFC and OC populations can test a possible radial gradient in the chemistry of icy planetesimals in the protoplanetary disk, along with the dynamical models that predict their dissemination. We report here the first quantitative comparison of three comets in Halley-type orbits.

Chemical diversity (e.g., for CO and hydrocarbon content) is reported among the small number of OC comets measured so far (Mumma et al. 2003, Biver et al. 2002); similar diversity is reported for $H_2S$ (Biver et al. 2006). The "organics-normal" group shows remarkable similarities to the composition of dense interstellar clouds and suggests processing of pre-cometary material at low (30 K) temperatures. However, the "organics-depleted" group differs greatly (Biver et al.

________________________________________________________________________

2006, Mumma et al. 2001, Bockelée-Morvan et al. 2001). The disrupted JFC comet 73P/Schwassmann-Wachmann 3 revealed a common composition for fragments B and C, with depleted organics akin to that in the OC comet C/1999 S4 LINEAR (Kobayashi et al. 2007, Dello Russo et al. 2007, Villanueva et al. 2006, Mumma et al. 2001). The controlled Deep Impact event (A'Hearn et al. 2005) revealed that material ejected from the nucleus of 9P/Tempel 1 (a JFC) was similar in composition to organics-normal OC comets, even though material released during the quiescent phase was organics-depleted (e.g., depleted in ethane, Mumma et al. 2005, DiSanti et al. 2007a). Mumma et al. (2005) argued that this difference could be interpreted in two ways: (1) internal heterogeneity resulting from radial migration of dissimilar planetesimals before aggregation, or (2) depletion of the surface layer by solar heating during prior apparitions.

Periodic comet 8P/Tuttle (hereafter, 8P) was discovered independently by Tuttle in Cambridge/USA (on 4 Jan. 1858) and by Bruhns in Berlin/Germany (on 11 Jan. 1858) (Vsekhsvyatskii 1964). An orbital period of 13.6 years (semi-major axis "a" = 5.67 AU), an inclination of about 55 deg and a Tisserand parameter $T_J < 2$ identifies the orbit of 8P as "Halley-type" (defined as $T_J < 2$ and "a" < 40 AU). This population was proposed to derive from the hypothesized inner Oort cloud, being captured into relatively short orbital periods primarily through gravitational interactions with the giant planets (Bailey and Emel'yanenko 1996, Emel'yanenko 1998). However, detailed dynamical models with a flattened inner Oort cloud (Levison, Dones, and Duncan 2001) failed to reproduce the observed orbital inclinations of the HTC population.

_______________________________________________________________________

Recent dynamical models (Levison et al. 2006) do reproduce the orbital distribution if HTCs originated from the scattered Kuiper disk (SKD, Duncan et al. 2004), were ejected to an inner Oort cloud, and then later re-inserted into the inner planetary region.  Some also return as dynamically new comets, and later can be seen as "returning OC comets" (Levison 1996).  In any case, the lifetime of HTCs in the neighborhood of the planets is of the order of $10^6$ years, before they are either ejected from the planetary system or they impact the Sun (Dones et al. 1999).  Since the JFCs also come from the SKD, one might expect compositional similarities among the HTC, JFC, and LPC groups though fractional representations within them might differ.  The number of observed Halley-family comets is small compared to that of the Jupiter-family of comets or to Oort cloud comets, although their absolute number may be large.

Comet 8P/Tuttle has been observed on 13 returns to perihelion, but conditions for study were especially favorable during the 2007/2008 apparition.  8P came close to Earth (minimum distance 0.25 AU on 2 Jan. 2008; maximum visual brightness $M_v$ ~5.3) with perihelion passage at 1.02 AU a few weeks thereafter on 27 Jan. 2008.  These conditions are highly favorable for infrared (IR) detections of parent volatiles from cometary ices and the direct measurement of abundance ratios in the sublimated gases.  Such opportunities are rare because they require the comet to be bright and also well placed in the heavens for study by large telescopes with specialized instruments.  While other comets in Halley-type orbits (109P/Swift-Tuttle, 27P/Brorsen-Metcalf, 96P/Machholz) have been targeted, native ice compositions were measured in comparable detail only for 1P/Halley and 153P/Ikeya-Zhang.

We measured the volatile composition of 8P using the high-resolution CRyogenic InfraRed

_______________________________________________________________________

Echelle Spectrograph (CRIRES) at the 8.2m Antu telescope of the Very Large Telescope Observatory (VLT; http://www.eso.org/paranal/) at Cerro Paranal in Chile. CRIRES (http://www.eso.org/instruments/crires/) provides long-slit spectra with extremely high spectral resolving power, approaching $\lambda/\Delta\lambda \sim 100,000$ using the 0.2″-wide slit. The observations were performed using the adaptive optics (AO) system (MACAO, http://www.eso.org/projects/aot/macao_vlti/), thereby minimizing slit-losses and achieving increased signal-to-noise in the central part of the coma as well as high along-slit spatial resolution, close to the diffraction limit of the telescope.

Observations of 8P and flux calibration standard stars were performed in service mode near perihelion (UT 27.02 Jan. 2008), spanning the UT dates 26 – 29 January 2008. A slit width of 0.4″ (resulting in $\lambda/\Delta\lambda \sim 50,000$) was used for all wavelength settings reported here. The methanol setting on 29 Jan. 2008 was additionally taken with the 0.2″ slit in order to achieve the highest possible wavelength resolution. In order to keep the comet on the slit, the MACAO AO system was locked on the central brightness condensation in the coma. This, coupled with differential auto-guiding of the telescope relative to a nearby star, allowed the comet to be tracked with very high reliability. Calibration spectra of the flux standard star were corrected for slit losses, using the spatial profile measured along the slit direction.

We used various instrument settings to sample wavelengths in the range ~ 2.9 - 4.7 μm. These settings encompassed emission lines of multiple parent volatiles, specifically $H_2O$, OH* (prompt emission), $CH_4$, $CH_3OH$, $C_2H_6$, CO and $H_2CO$. For all observations, the telescope was nodded along the slit in an ABBA sequence with the two beams separated by 15″ (one-half the length of

_________________________________________________________________________

the slit). This provided cancellation (to second order in air mass) of thermal background emission and of line emission from the terrestrial atmosphere, through the arithmetic operation (A – B – B + A). In addition to this standard nodding, the comet was "jittered" along the slit by adding a small random (but well known) offset of up to ~ ± 2″. The jittering process provided protection against the object (comet or star) falling repeatedly on bad pixels or other imperfections in the detector array. The echellograms were processed using custom algorithms tailored for CRIRES (including jitter correction); these are based on well-tested data reduction and analysis routines developed for the NIRSPEC and CSHELL instruments (Villanueva et al. 2008, Bonev 2005, DiSanti et al. 2001). Frequency calibration and correction for telluric absorption utilized synthetic transmittance spectra based on rigorous line-by-line, layer-by-layer radiative transfer modeling of the terrestrial atmosphere using GENLN2 (Edwards 1992, Hewagama et al. 2003).

For each setting, a modeled continuum for the dust (multiplied by the atmospheric transmittance) was subtracted from the spectrum measured for 8P, thereby isolating the molecular emission lines (see Fig. 1). Production rates (Q) were retrieved through comparison of measured line fluxes and corresponding fluorescence efficiencies (g-factors) at the appropriate (measured) rotational temperature ($T_{rot}$). We measured nineteen $H_2O$ lines (e.g., Fig. 1A - 1D) that together span a wide range of rotational energies, enabling us to determine $T_{rot}$ by means of rigorous correlation and excitation analyses (DiSanti et al. 2006, Bonev 2005). This measured $T_{rot}$ (60 +8/-9 K, see Table 1) was adopted for other parent volatiles.

The g-factors are based on quantum-mechanical fluorescence models for $H_2CO$ (DiSanti et al

2006, Reuter et al. 1989), $H_2O$ (Dello Russo et al. 2004), $C_2H_6$ (Dello Russo et al. 2001), $CH_4$ (Gibb et al. 2003), $CH_3OH$ (3.5 μm, Reuter et al. 1992) and CO (DiSanti et al. 2001), and on empirical efficiencies for OH* (Bonev et al. 2006) and $CH_3OH$ (3.3 μm). OH* can be used as a proxy for $H_2O$ production when direct $H_2O$ measurements are not available. We verified that our reported production rates are only weakly influenced by the uncertainty in rotational temperature. The uncertainties in Q given in Table 1 reflect the level of agreement among production rates obtained independently from individual lines of a given species. Errors in mixing ratios also incorporate uncertainties in comparing results from different settings when water was not measured simultaneously with the trace species.

In the spectra presented here, we detected five parent volatile species: $H_2O$ (e.g. Fig. 1A-1D), $CH_4$ (and OH*, Fig. 1E), $C_2H_6$ (and $CH_3OH$, Fig. 1F), CO (and $H_2O$, Fig. 1G and 1H), and $CH_3OH$ (and OH*, Fig. 1I). An upper limit for $H_2CO$ was also obtained (Fig. 1I). The extracted production rates and mixing ratios for 8P/Tuttle are listed in Table 1. The composition of 8P is compared with two other comets in Halley-type orbits and also with the mean of five returning OC comets in Table 2.

In 8P, the abundance ratio for methanol (3.3%) is higher by almost a factor of two than the value for 1P/Halley and than the mean (~2%) of five "organics-normal" OC comets and, and it is marginally higher than in 153P/Ikeya-Zhang (Table 2). CO in 8P (0.45%) falls well below the low end of the range (1.8% – 17%) of the five OC comets, and is much lower than in either 1P/Halley or 153P/Ikeya-Zhang. $CH_3OH$ is much lower in the organics-depleted comets C/1999 S4 LINEAR (< 0.15%, 2-σ) and 73P/Schwassmann-Wachmann 3 (< 0.3%, 2-σ, Villanueva et al.

___________________________________________________________________________

2006, and even as low as ~ 0.1% - 0.3% Dello Russo et al. 2007), but CO in these organics-depleted comets agrees with 8P (Mumma et al. 2001, DiSanti et al. 2007b). $CH_4$ (0.37%) falls within the range (0.5%-1.5%) of the five OC comets, and agrees with the upper limit in 1P/Halley. $C_2H_6$ in 8P (0.30%) is lower than the mean (0.6%) of the five OC comets (Table 2) by approximately a factor of two, is comparably lower than in 153P/Ikeya-Zhang and is marginally lower than the abundance in 1P/Halley. Taken at face value, the hydrocarbon chemistries of comets 8P/Tuttle, 1P/Halley, and 153P/Ikeya-Zhang do not reveal any shared characteristics that might distinguish comets in Halley-type orbits from other comets of OC origin in this sample.

Several unusual properties have been reported for 8P/Tuttle. Radar images (on Jan. 2-4) show a strongly bifurcated nucleus, possibly a contact binary, with two roughly spherical lobes measuring 3 and 4 km in diameter (± 25 percent) (Harmon et al. 2008). Their preliminary estimate for the rotation period is 7.7 ± 0.2 hr. Velocity profiles for HCN in 8P (Dec. 29-Jan. 2) are consistent with rotation periods of 5.7 hr, 7.4-7.6 hr, and multiples thereof (Drahus et al. 2008). The line area remained constant to a precision of about 30 percent, suggesting a constant production rate for HCN at that precision. Optical jets implied a rotation period of 5.71 ± 0.04 hr (Dec. 15-16); the coma morphology was consistent with a single source (Schleicher and Woodney, 2008). Schleicher (2008) found that 8P has "typical" composition and a very low gas-to-dust ratio, based on optical measurements of "daughter" molecules (OH, CN, $C_2$, $C_3$, and NH), in agreement with earlier results (A'Hearn et al. 1995). All three "Halley-type" comets discussed here are ranked as "typical" based upon the measured daughter products, but strong differences in composition are seen when abundance ratios of parent volatiles are examined (Table 2).

Unedited final draft, to appear in: ApJ 683:L71-L75 (August 10, 2008).
______________________________________________________________________

We thank the VLT science operations team of the European Southern Observatory for efficient execution of the observations. This work was supported by the German-Israeli Foundation for Scientific Research and Development under grant I-859-25.7/2005, the International Max-Planck Research School, and NASA's Planetary Astronomy Program, Astrobiology Program, and Postdoctoral Program.

______________________________________________________________________

___________________________________________________________________

**Table 1.** The parent volatile composition of comet 8P/Tuttle as measured with CRIRES/VLT.

| Species | Date/Time | Trot [K] | Mean ν [cm$^{-1}$] | Lines sampled | Q [$10^{26}$ s$^{-1}$] | Mixing ratio [%] |
|---|---|---|---|---|---|---|
| **UT 26 January 2008** | | | | | | |
| $C_2H_6$ | 01:40 to 02:18[1] | (60) | 2988 | 5 | 1.77 ± 0.18 | 0.30 ± 0.03 |
| $CH_4$ | " | (60) | 3014 | 2 | 2.12 ± 0.50 | 0.36 ± 0.09 |
| **UT 27 January 2008** | | | | | | |
| $H_2O$ | 01:28 to 02:06[2] | 60 +8/-9 | 3445 | 19 | 597 ± 27 | 100 |
| $C_2H_6$ | 02:13 to 02:52[3] | (60) | 3000 | 1 | 1.68 ± 0.36 | 0.28 ± 0.06 |
| $CH_4$ | " | (60) | 3012 | 3 | 2.20 ± 0.47 | 0.37 ± 0.08 |
| $CH_3OH$ | " | (60) | 2999 | 4 | 20.0 ± 1.90 | 3.36 ± 0.40 |
| $H_2O$ | 02:59 to 03:53[4] | (60) | 2160 | 3 | 544 ± 38 | 100.0 |
| CO | " | (60) | 2156 | 11 | 2.43 ± 0.43 | 0.45 ± 0.09 |
| **UT 29 January 2008** | | | | | | |
| $CH_3OH$ | 01:14 to 02:00[5] | (60) | 2844 | Q-branch | 19.4 ± 1.61 | 3.24 ± 0.32 |
| $H_2CO$ | " | (60) | 2834 | 9 | < 0.5 | < 0.1 |

The cometary ephemerides (heliocentric distance [AU], geocentric distance[AU], geocentric velocity [km s$^{-1}$]) during the observations were: 1 (1.027, 0.494, 24.73); 2 (1.027, 0.509, 24.76); 3 (1.027, 0.509, 24.80); 4 (1.027, 0.509, 24.84); 5 (1.028, 0.537, 24.76). All mixing ratios are based on the high-quality water production rate measured on 27 Jan 01:28 - 02:06 UT, excepting CO for which $H_2O$ was measured simultaneously. The formaldehyde value represents the 3σ upper limit, based on stochastic error.

**Table 2.** The native parent volatile composition of three comets in Halley-type orbits.

| Comet | CO | $CH_4$ | $C_2H_2$ | $C_2H_6$ | $CH_3OH$ | $H_2CO$ |
|---|---|---|---|---|---|---|
| 8P/Tuttle[a] | 0.45 ± 0.09 | 0.37 ± 0.06 | NS | 0.30 ± 0.03 | 3.3 ± 0.26 | < 0.1 |
| 1P/Halley[b] | 3.5 | < 1 | ~ 0.3 | ~ 0.4 | 1.7 ± 0.4 | < 0.4 |
| 153P/Ikeya-Zhang[b] | 4.7 ± 0.8 | 0.51 ± 0.06 | 0.18 ± 0.05 | 0.62 ± 0.13 | 2.5 ± 0.5 | 0.62 ± 0.18 |
| 5 OC Comets[b] | 1.8 - 17 | 0.5 – 1.5 | 0.2 – 0.3 | 0.6 | 2 | IP |
| C/1999 S4 LINEAR[b] | 0.9 ± 0.3 | 0.18 ± 0.06 | < 0.12 | 0.11 ± 0.02 | < 0.15 | IP |

[a]This work. Orbital parameters (a, e, i, $T_J$) are: (17.834, 0.96714, 162.263, -0.605) for 1P, and (51.275, 0.99011, 28.120, 0.878) for 153P.
[b]Mean or range. Original citations are given in Mumma et al. (2003). NS = not sampled. IP = In Progress.

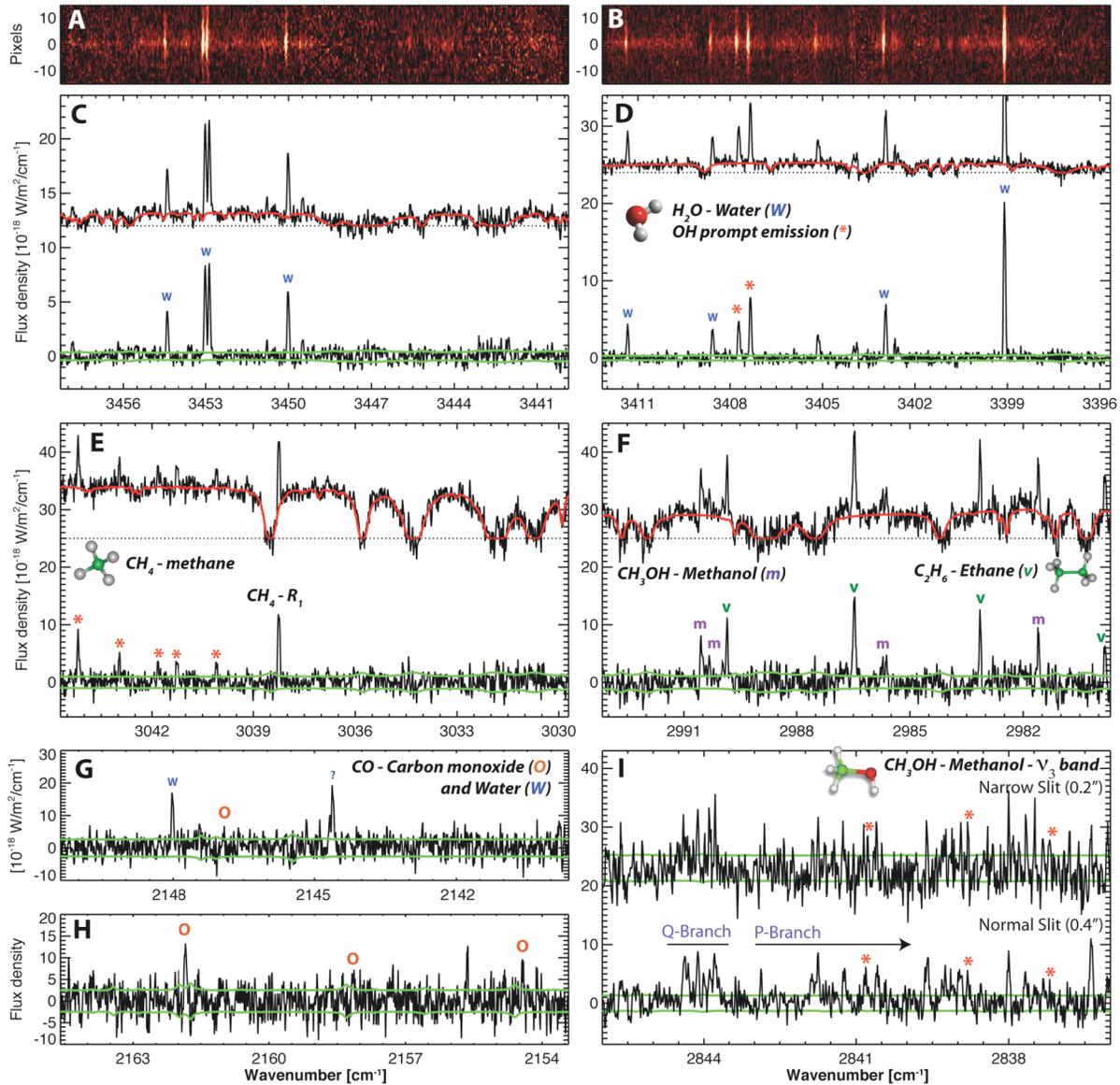

**Figure 1**: Detection of parent volatiles and dust in comet 8P/Tuttle in January 2008. The thick red lines in panels C-F represent the cometary continuum convolved with a synthetic transmittance spectrum of the terrestrial atmosphere, and the green lines in panels C-I indicate the noise envelope ($\pm$ 1-$\sigma$). Panels A and B show echellograms for CRIRES detectors #2 and #4 of the 2.9 $\mu$m wavelength setting; and the corresponding spectral extracts ($\pm$7 rows from the nucleus) are presented in panels C and D. Lines of $H_2O$ (w) appear in panels C and D, and OH* prompt emission (*) is seen in D and E. Panels E and F show detections of methane ($CH_4$ - R1), ethane (v), methanol (m) and OH* prompt emission (*) measured near 3.3 $\mu$m. Detections of water (w) and carbon monoxide (o) near 4.7 $\mu$m are shown in panels G and H. The $\nu_3$ band of methanol and OH* prompt emission appear in panel I. All detections were acquired with the 0.4″ slit, but methanol was also taken with the narrowest slit (0.2″) to resolve this complex band system. Quantitative results are given in Table 1.